\documentclass[12pt]{iopart}
\pdfoutput=1
\usepackage{iopams}
\usepackage{graphicx}

\begin{document}

\title[Quantum noise with exchange and tunneling.]{Quantum noise with exchange and tunneling: Predictions for a two-particle scattering experiment with time-dependent oscillatory potentials.}

\author{E. Colom\'{e}s, D. Marian  and  X. Oriols}

\address{Departament d\rq{}Enginyeria Electr\`{o}nica, Universitat Aut\`{o}noma de Barcelona, Spain.}
\ead{xavier.oriols@uab.es}

\begin{abstract}

Quantum noise with exchange and tunneling is studied within time-dependent wave packets. A novel expression for the quantum noise of two identical particles injected simultaneously from opposite sides of a tunneling barrier is presented. Such quantum noise expression provides a physical (non-spurious) explanation for the experimental detection of two electrons at the same side under static potentials. Numerical simulations of the two-particle scattering probabilities in a double barrier potential with an oscillatory well are performed. The dependence of the quantum noise on the electron energy and oscillatory frequency are analyzed. The peculiar behaviour of the dependence of the quantum noise on such parameters is proposed as a test about the soundness of this novel quantum noise expression, for either static or oscillatory potentials.

\end{abstract}

\maketitle

\section{Introduction}
\label{sec1}

In general, the own definition of transport implies movement, variations in time. However, because solving time-dependent transport models entails a large computational burden \cite{xavier1}, there are many examples of time-independent models that provide accurate predictions about transport phenomena. Specifically, in quantum transport, a normal and extended approximation is to substitute the intrinsic time-dependence of the states involved in the computations by time-independent ones \cite{Buttiker2}-\cite{datta}. The success of such time-independent models to solve in a comfortable way many quantum transport phenomena is unquestionable. However, can we always remove the explicit time-dependence of transport models? \cite{Pastawski}-\cite{Jauho} The answer is not simple at all.\\

\begin{figure}[h]
\begin{center}
\includegraphics[scale=1.2]{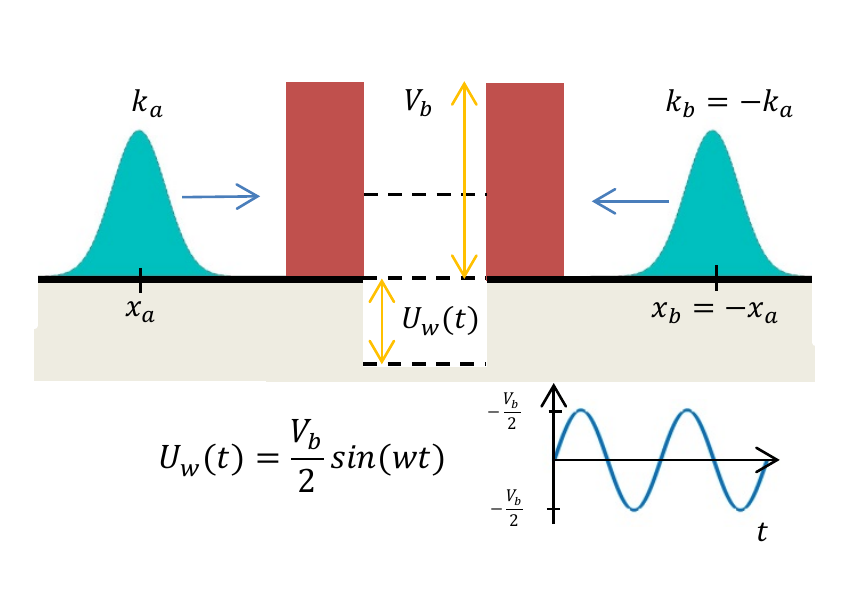}
\caption{Double barrier with a time-dependent oscillatory quantum well $U_w(t)$, we see its evolution in the inset figure. With this experiment, we can prove the reliability of our predictions. Two wave packets ($\phi_{a}(x)$ and $\phi_{b}(x)$) are located at each side of the barrier at the same distance ($x_b = -x_a$) describing each one an electron with the same energy but with opposite momentum $k_b=-k_a$.}
\label{fig1}       
\end{center}
\end{figure}

In this paper, we analyze if such time-dependence of the states is relevant or not in a type of Hong-Ou-Mandel (HOM) experiment \cite{Boc} with tunneling and exchange. Two identical electrons are injected simultaneously from two different inputs and after scattering on an electron beam splitter they are measured at two different outputs. The correlation between the detection of the two outputs is measured depending on the injection delay. From these correlation values one can directly obtain quantum noise, i.e. the fluctuations in the number of detected electrons. In particular, we focus on a situation quite close to the experiment mentioned above, but where the scattering region is a double barrier potential with an oscillatory quantum well in a 1D system, see \fref{fig1}. In particular, we focus on the case where there is no delay in the injection among both electrons. Then, in principle it is expected that quantum noise is suppressed due to Pauli principle, which states that two electrons cannot be at the same place with the same state \cite{Pauli}. As a consequence, it is expected that each electron will be located at a different output with no (zero frequency) fluctuations. However, our numerical results and the experiment in this type of HOM system show that, even if quantum noise is reduced it is not completely suppressed, indicating the non-zero probability of detecting simultaneously two electrons at the same side.\\

We have shown in a previous paper \cite{arxiv} that, because of the localization of the initial states and the energy dependence of the scattering region, the probability of having two electrons at the same place is different from zero, even when the (Fermi or Bose) nature of particles is explicitly considered. In this paper we develop a novel expression of the quantum noise in this two-particle scattering process, which takes into account these unexpected probabilities. This expression gives a fundamental unavoidable reason for the quantum noise unexpected enhancement \cite{arxiv}. 
Finally, with the situation described in \fref{fig1}, similar to the HOM one \cite{Boc}, we propose an experiment with oscillatory potentials that is able to test the signature of our contribution over other possible explanations of the unexpected experimental results in this type of HOM systems (decoherence \cite{feve}, spurious effects \cite{Tarucha}, time delay \cite{MOSKALETS} or interaction between different modes \cite{placais}).\\

The paper is organized as follows. In \Sref{sec2} we discuss the two-particle scattering and how it is possible to measure at the same position two initially identical electrons after the interaction with the time-dependent potential barrier. In \Sref{sec3a} we explain how quantum noise is computed taking into account these new two-particles scattering possibilities. We named these probabilities as ``new'' because a zero probability for such process is predicted by the Landauer-B\"uttiker formalism \cite{Landauer1,Buttiker4}. In \Sref{sec3} we explain a procedure which will be able to test the soundness of our proposal, through a time-dependent oscillatory double barrier potential. Finally, in \Sref{sec4} we conclude.

\section{Two-particle scattering}
\label{sec2}

\subsection{The system}

\begin{figure}[h]
\includegraphics[scale=0.43]{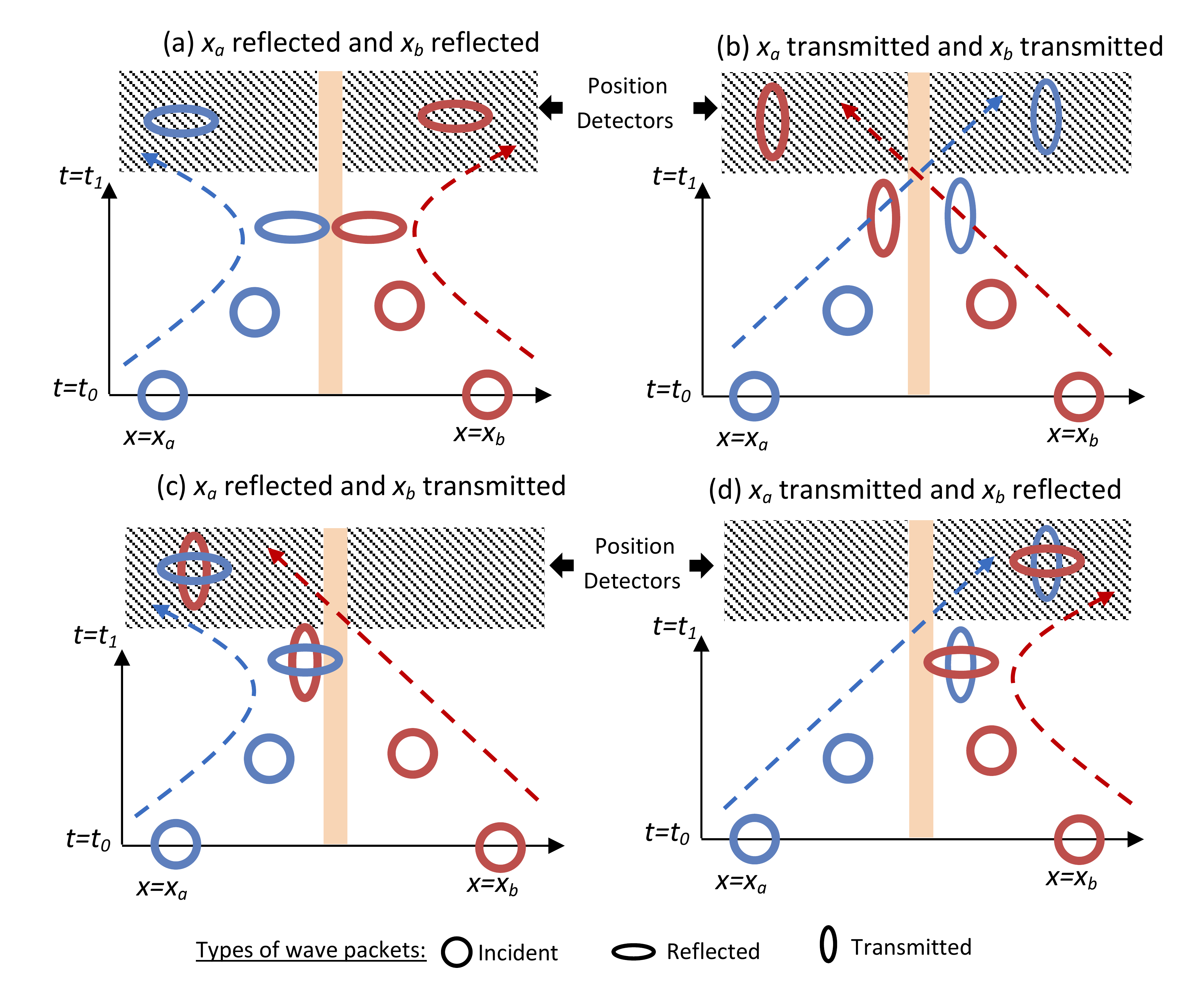}
\caption{Two identically injected wave packets from the left $x_a$ and from the right $x_b$ of a scattering barrier. Solid regions represent the barrier region and shaded regions represent the particle detectors. (a) and (b) each particle is detected on a different side of the barrier  at final time $t_1$ when the interaction with the barrier has almost finished. (c) and (d) both particles are detected on the same side of the barrier.}
\label{fig2}       
\end{figure}

We consider a oscillatory potential barrier system with two identical electrons, initially located at each side of the barrier at the same side, but with opposite momentum (\fref{fig2}). We solve the two-particle time-dependent Schr\"odinger equation,

\begin{equation}
i \hbar \frac{\partial \Phi}{\partial t} = \left[ - \frac{\hbar^2}{2m^*}\frac{\partial^2}{\partial x_1^2} - \frac{\hbar^2}{2m^*}\frac{\partial^2}{\partial x_2^2} + V(x_1,t)+V(x_2,t) \right] \Phi,
\label{scho}
\end{equation} 

where $m^*$ is the effective electron mass and $V(x_i,t)$ takes into account the one-particle interaction between one electron and the time-dependent tunneling barrier depicted in \fref{fig1}. The exchange interaction is introduced in equation \eref{scho} in the shape of the initial wave function $\Phi(x_1,x_2,t_0)$. The anti-symmetrical many-particle wave function for two electrons is:

\begin{equation}
\Phi(x_1,x_2,t_0)=\frac {\phi_a(x_1,t_0)\phi_b(x_2,t_0) - \phi_a(x_2,t_0)\phi_b(x_1,t_0)} {\sqrt{ 2}}.  
\label{initial}
\end{equation}

After the interaction with the barrier, in addition to the usual scattering probabilities where both electrons are found at each side of the barrier (\fref{fig2} a) and b)), it is also possible to find both of them at the same side of the barrier, i.e. both at the left or both at the right side (\fref{fig2} c) and d)).

The only requirement to get the novel probabilities is that the initial states ($\phi_a$ and $\phi_b$) are described by localized (normalizable) wave functions (and not the scattering states \cite{Landauer1,Buttiker4}, which are infinitely extended in the space) and an energy dependence in the scattering region. These new final scenarios are not in contradiction with the Pauli principle as mentioned in the introduction. In our case, they can occur because, after the interaction with the barrier, the transmitted and reflected wave functions of the electrons suffer a different evolution and therefore they do not overlap, or at least not completely, allowing the possibility of being at the same place (\fref{fig2} c) and d)) with the same energy, but different spatial shape. For further discussion, see Ref. \cite{arxiv}.

\subsection{The new probabilities}

These new probabilities are reflected in equations \eref{P4b} and \eref{P3b} (proved and derived in Ref. \cite{arxiv}). The former describes the probability of two electrons being, both, at the left side of the barrier:

\begin{eqnarray}
\mathcal{P_{LL}}=\int^{0}_{-\infty} dx_1 \int^{0}_{-\infty} dx_2 \;\; |\Phi |^2=R_a T_b - |I^{r,t}_{a,b}|^2,
\label{P4b}
\end{eqnarray}

the latter  provides the probability of both electrons being located at the right side:

\begin{eqnarray}
\mathcal{P_{RR}}=T_a R_b - |I^{r,t}_{a,b}|^2.
\label{P3b}
\end{eqnarray}

Finally,  the probability of one particle at each side is:

\begin{eqnarray}
\mathcal{P_{LR}}=R_a R_b +T_a T_b + 2 |I^{r,t}_{a,b}|^2.
\label{P1b}
\end{eqnarray}

In these expressions, $R_i$ and $T_i$ are the single-particle reflection and transmission coefficient of the i-wave packet. $I^{r,t}_{a,b}$ is the important overlapping term among the different wave packets after the interaction with the barrier:

\begin{eqnarray}
I^{r,t}_{a,b}=\int_{-\infty}^{0} dx \; \phi_a^r(x,t_1)  \; \phi_{\;b}^{*t}(x,t_1), 
\label{integral} 
\end{eqnarray}

where $\phi_a^r$ and $\phi_b^t$ are the reflected component of the initial state $\phi_a$ and the transmitted component of the initial state  $\phi_b$ respectively. The time $t_1$ can be any time large enough so that the probability presence in the barrier region remains negligible. 

Depending on the contribution of the overlapping term in equation \eref{integral}, the probabilities \eref{P4b}, \eref{P3b} and \eref{P1b} can achieve two different particular limits:

\begin{itemize}
\item The Landauer-B\"uttiker results \cite{Buttiker4} are recovered when $|I^{r,t}_{a,b}|^2$ is equal to $RT$ and therefore $\mathcal{P_{RR}} =\mathcal{P_{LL}} =0$, and  $\mathcal{P_{LR}} =1$. Thus there is no possibility of finding both particles at the same place. This limit is achieved when the wave packets are spatially large and similar (not identical) to an infinitely extended scattering states.

\item The results for distinguishable particle are obtained when $I^{r,t}_{a,b}=0$, i.e. the reflected wave packet and the transmitted one are orthogonal. Then, $\mathcal{P_{RR}}=T_a R_b$, $\mathcal{P_{RR}}=T_a R_b$ and $\mathcal{P_{LR}}=R_a R_b +T_a T_b$. This case is achieved when the wave packets, after the interaction with the barrier, are very different. For example, this occurs in the case of a double barrier, when both electrons have the resonant energy, but one is transmitted and the other is reflected. 
\end{itemize}

\section{Quantum noise with the new probabilities}
\label{sec3a}

In this section, after having developed the new two-particle scattering probabilities $\mathcal{P_{RR}}$, $\mathcal{P_{LL}}$ and $\mathcal{P_{LR}}$, we compute the quantum noise formula with the new possibilities (\fref{fig2}  c) and d)) described above. The new formula will be compared to the Landauer-B\"uttiker  expression and the ``semiclassical" equation for noise without exchange.\\

For simplicity, we consider a symmetric system, where $T_a=T_b$, $R_a=R_b$ and thus $\mathcal{P_{RR}} =\mathcal{P_{LL}}$. Only one- and two-particle scattering processes are treated. The extension to many-particle processes will be detailed in a future work \cite{Future}. At low frequencies, when the displacement current is neglected, the noise can be computed from the knowledge of the number $N$ of transmitted particles through the barrier during the time $t_d$:

\begin{equation}
\langle S \rangle= lim_{t_d \rightarrow \infty} 2 q^2\frac{\langle N^2 \rangle_{t_d}-\langle N \rangle_{t_d}^2}{t_d}.
\label{S}
\end{equation}

\begin{figure}[h]
\centering
\includegraphics[scale=0.4]{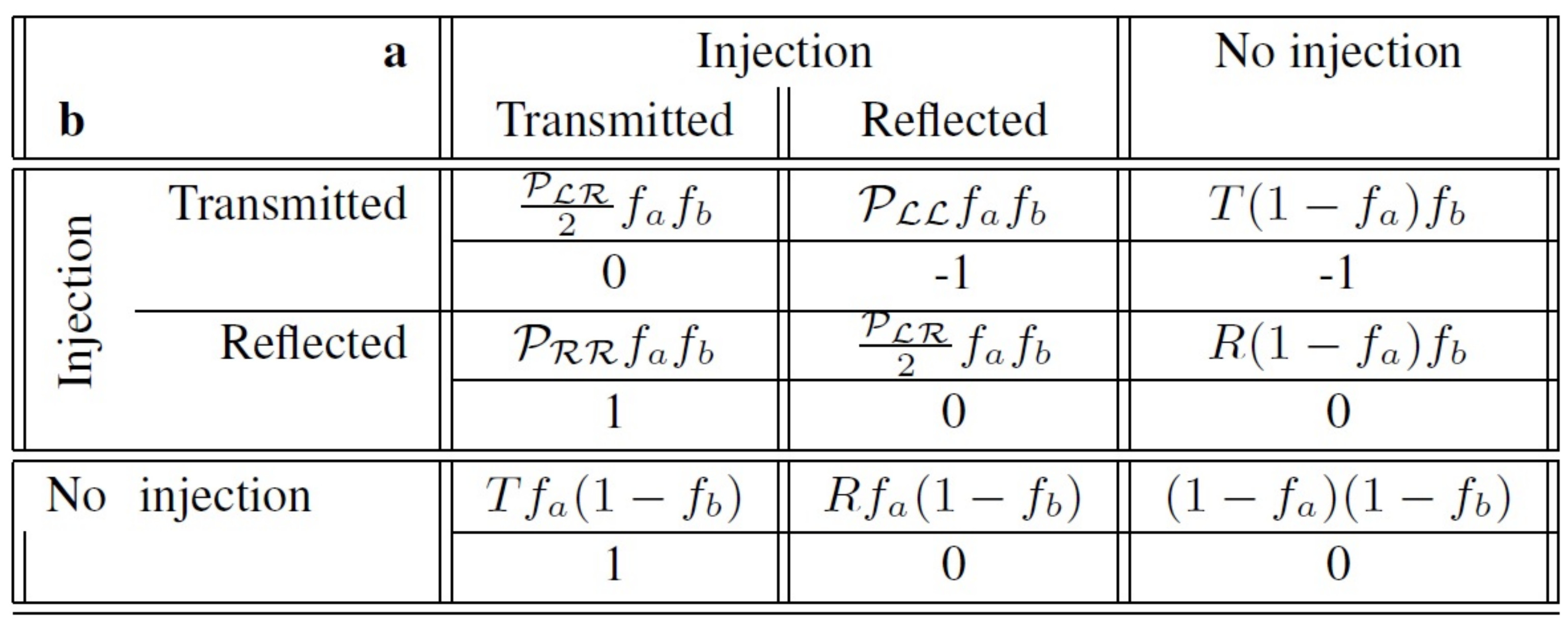}
\caption{Probability (upper) that $N$ (lower) electrons are transmitted from the left to right reservoir during the time interval $t_d$. $f_i$  is the Fermi distribution ($i=a,b$) and $T_i$ and $R_i$ the $i$-wave packet single particle transmission and reflection coefficients.}
\label{table2}
\end{figure}

We define $\langle N \rangle_{t_d} = \sum_{N=-\infty}^{N=\infty} P(N) N$  and $\langle N^2 \rangle_{t_d} = \sum_{N=-\infty}^{N=\infty} P(N) N^2$, where $P(N)$ is the probability of $N$ particles being transmitted from the left to the right reservoir. The probabilities  $P(N)$ are computed from the direct solution of the two-particle Schr\"odinger  equation including exchange interaction (equation \eref{scho}) and summarized in \fref{table2}.

Now, one can determine the noise $\langle S \rangle$, which due to the new  $P(N)$, is related to $\mathcal{P_{LL}}$ and $\mathcal{P_{RR}}$:

\begin{eqnarray}
\langle S \rangle \!\!=\!\! \frac{4q^2}{h}(T [f_a(1-f_a)  +f_b(1-f_b)] \!+\! T(1-T)(f_a-f_b)^2\!+\!2\mathcal{P_{LL}}f_af_b).
\label{S2}
\end{eqnarray}

Expression \eref{S2} contains the usual Landauer-B\"uttiker formalism noise expression in the case where $\mathcal{P_{RR}}=\mathcal{P_{LL}}=0$. However, we have showed that generally, $\mathcal{P_{LL}}\neq 0$ and quantum noise is increased. In the limit of the distinguishable particles behaviour, the classical noise results are recovered $\mathcal{P_{LL}}= RT$ and electrons behave as classical particles, without exchange interaction. In general, the results predicted by equation \eref{S2} lie among the Landauer-B\"uttiker formalism and the classical results.\\

\section{Numerical results for the oscillatory proposed experiment}

\label{sec3}

As mentioned in \sref{sec1}, there is a HOM experiment \cite{Boc} where two identical electrons are injected simultaneously from two inputs and measured in two outputs. In the experiment, it is found out that the possibility of measuring both electrons at the same side is not zero, as usually expected. Apart from F\`eve \textit{et al} \cite{placais} that explain this result because of the interaction among different Landau levels in the inner channels when injecting the electrons, other explanations appeal for decoherence \cite{feve}, spurious results \cite{Tarucha} and time delay in the injection \cite{MOSKALETS}.\\

With the approach explained previously in \sref{sec2}, alternatively, a fundamental reason is given for the experiment results. Due to the time- and energy-dependence evolution of the electron wave packet, after the interaction with the barrier, the reflected and transmitted components of the wave packets do not overlap completely. Thus, there is no reason to expect that they cannot be detected at the same place (according to equations \eref{P4b} and \eref{P3b}) because their states are different and then the exclusion Pauli principle does not apply. The reader can find more details in Ref. \cite{arxiv}.\\

As it has been already exposed, one purpose of this work is to propose an experiment which would be able to test the reliability of our time-dependent explanation for the unexpected noise results. We analyze the case where two electrons are injected simultaneously with the same energy from both sides of a double barrier at the same distance from the barrier. This double barrier system has a time-dependent well (see \fref{fig1}), which oscillates periodically according to expression $U_w=\frac{V_b}{2}sin(wt)$. In order to increase the visualization of these new probabilities, we will consider the injection of electrons whose energies are close to the (first) resonant energy of the double barrier. For these energies, the transmission coefficient has a sharp energy-dependence so that the reflected and transmitted wave packets become almost orthogonal, $|I^{r,t}_{a,b}|^2 \approx 0$. See further details in Ref. \cite{arxiv}. Then, the new probabilities \eref{P4b} and \eref{P3b} become more relevant.\\

We expect that, by changing the potential level of the well inside the barrier, the resonant energy of the double barrier will change accordingly.  This will cause that for some electrons that in the time-independent case were not resonant, and therefore their probability of finding both at same place was low, will be resonant, increasing enormously the probabilities of finding both of them at the same side of the barrier.\\

In our proposed experiment, there is always two electrons, one injected from each side of the barrier simultaneously, and therefore the only term which survives in equation \eref{S2} is the last one (because $f_a=f_b=1$), which contains the new probability $\mathcal{P_{LL}}$ of finding both particles at the same side. Therefore, the computation of $\mathcal{P_{LL}}$ provides directly, apart from a constant factor, the quantum noise in equation \eref{S2}.\\

\begin{figure}[h]
\centering
\includegraphics[scale=0.6]{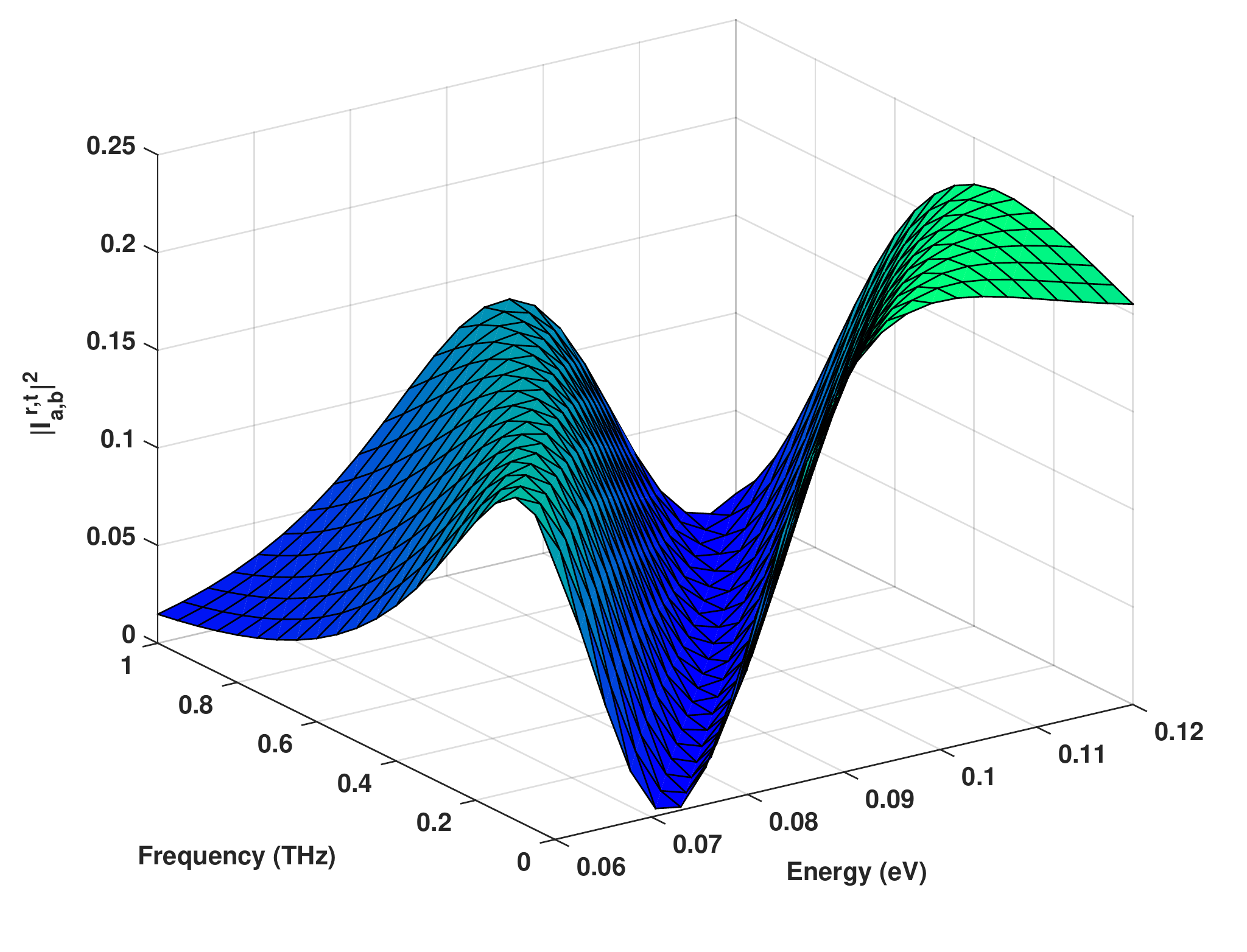}
\caption{The overlapping term $|I^{r,t}_{a,b}|^2$ is plotted as a function of the frequency of the oscillation well and also as a function of the central energy of the injected electrons. We see that for certain values, the overlapping is almost zero, corresponding to the resonant energies.} 
\label{ITD}
\end{figure}

\begin{figure}[h]
\centering
\includegraphics[scale=0.6]{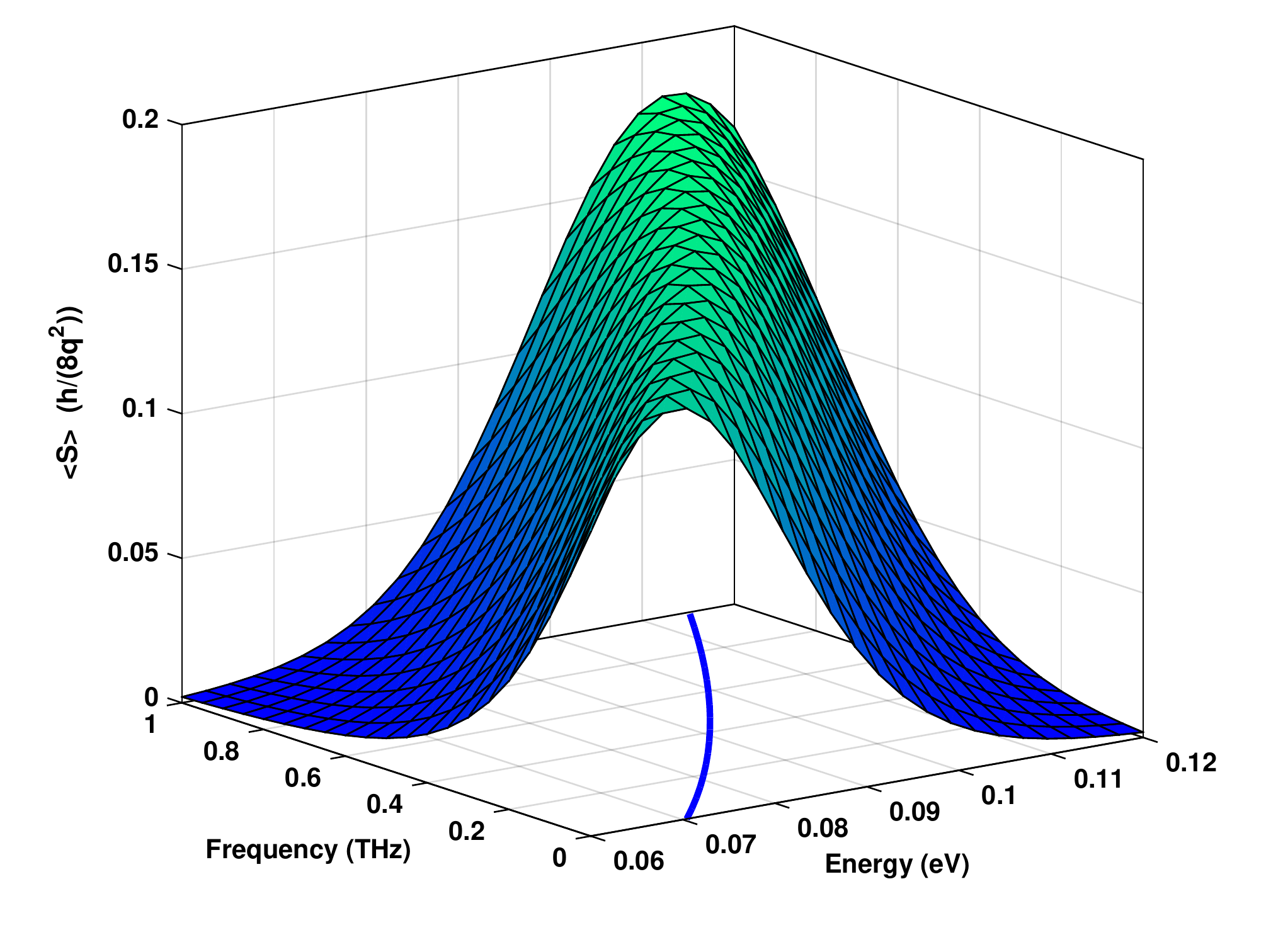}
\caption{Noise is plotted as a function of the frequency of the oscillation well and also as a function of the central energy of the injected electrons. We appreciate how there is are a line where maximum values are achieved. In the blue solid line, we appreciate the maximum values expected for the noise from equation \eref{ww}, the data fit accurately.} 
\label{PLLTD}
\end{figure}

We performed simulations for the experiment, as a function of the energy of the injected electrons and also of the frequency of the oscillation of the bottom potential of the quantum well. We chose as initial state for the electron wave functions $\phi_a(x,t_0)$ and $\phi_b(x,t_0)$ a Gaussian function $\phi_i=[2/(\sigma^2\pi)]^{1/4} e^{ik_0(x-x_0)}e^{-(x-x_0)^2/\sigma^2}$, whose initial position is $x_0=175\:nm$ far from the center of the barrier, dispersion $\sigma=50\:nm$ and initial central momentum is $k_0=\sqrt{2m^*E}/\hbar$. The barrier is $0.4\:eV$ high, its thickness is $1.0\:nm$ and the quantum well length is $5.2\:nm$. We emphasize that any other localized wave function can be chosen without modifying qualitatively the results discussed in this work.\\

This result is seen in figures \ref{ITD} and \ref{PLLTD}, where we plot the overlapping term $I^{r,t}_{a,b}$ (equation \eref{integral} and noise $\langle S \rangle$ (equation \eref{S2}) respectively, as a function of the energy of the electrons and also of the oscillatory frequency of the well. The results corresponding to the static case (no oscillatory well) are seen at frequency equal to zero. In this static situation, the resonant energy is $E_r=0.073\:eV$, and we appreciate in \fref{ITD} that at this energy value, the overlapping is minimum, and in \fref{PLLTD} that noise achieves its maximum value.\\

In figure \ref{ITD} (figure \ref{PLLTD}), we observe that when we switch on the oscillation, the minimum (maximum) value for the overlapping $I^{r,t}_{a,b}$ (noise $\langle S \rangle$) moves. Therefore, as we expected, resonant energies could be found for other energies even if in the static case they were not. Moreover,  the behaviour of the movement of the maximum values can be easily understood. As time passes by, the well potential increases. As frequencies are not very high (compared to the inverse of the transit time $1/\tau_t$), a first valid approximation to compute the resonant energy is: 

\begin{eqnarray}
E_r=E_{r0}+\frac{V_b}{2}sin(wt_b),
\label{er}
\end{eqnarray}

where $E_{r0}$ is the resonant energy when there is no oscillation in the well and $t_b$ the time that the electron takes to arrive to the barrier.  This time $t_b$ is the space from the place where the injection is carried out until the barrier ($x_0$), divided by the velocity of the electron ($v_e$):

\begin{eqnarray}
t_b=\frac{x_0}{v_e}=\frac{x_0 m^*}{\sqrt{(2m^*E_r)}}.
\label{tb}
\end{eqnarray}

From equations \eref{er} and \eref{tb} one realizes that the frequency for the maximum probabilities as a function of the resonant energy is:

\begin{eqnarray}
w(E_r)=\frac{\sqrt{(2m^*E_r)}}{x_0 m^*}arcsin\Big(\frac{2(E_r-E_{r0})}{V_b}\Big),
\label{ww}
\end{eqnarray}

which is in perfect agreement with the results observed in \fref{PLLTD}. There, we see that the peak values for noise at each energy, move accordingly to equation \eref{ww}, which is plotted in the $frequency$-$energy$ plane with a blue solid line.\\

Therefore, the simulations performed with sinusoidal potentials provide a clear behaviour: when we move to higher frequencies, quantum noise will be increased as we move to higher energies and will achieve a maximum at the new resonant energy. The experiment can be modified and include other behaviours when changing the potential according to another expression. For instance, $U_w=-\frac{V_b}{2}sin(wt)$, in this case the resonant energy will decrease as frequency increases (mathematically, the negative sign in front of the sinusoidal signal can also be introduced as a negative frequency in the plot \fref{PLLTD}). In this work, we propose that this very particular behaviour of the maximum of the quantum noise $\langle S \rangle_{max}$ in the $frequency$-$energy$ plane can be use as a test of our novel physical explanation of non-zero correlations in this type of HOM experiments with exchange and tunneling. The experimental confirmation of such predictions will, in fact, give support for the need of using time-dependent states when modelling quantum noise in such experiments even for static (DC) conditions.   

\section{Conclusions}
\label{sec4}

Motivated by the Hong-Ou-Mandel kind experiment on quantum noise performed with electrons \cite{Boc}, we analyze a similar two-particle scattering scenario with exchange and tunneling.  We inject two identical electrons with opposite momenta in a double barrier potential with an oscillatory well. Then, because of the different evolution suffered by the transmitted and reflected wave packet, we prove that two electrons can be found at the same place after the interaction with the barrier with probabilities given by equations \eref{P4b} and \eref{P3b}. These new probabilities lead to a novel quantum noise expression (equation \eref{S2}). We remark that this formula contains two particular and interesting limits: the quantum Landauer-B\"uttiker noise expression and the distinguishable classical noise result.\\

Finally, we perform numerical simulations for oscillatory potentials which can certify the soundness of the new probabilities explained above. We compute the new quantum noise as a function of the energy of the initial electrons and of the frequency of the oscillatory well. We show in \fref{PLLTD} that, the maximum value of noise changes in energy as frequency changes according to equation \eref{ww}. We propose to reproduce these type of HOM experiment with an oscillatory well. Then, the satisfactory test on the experimental agreement of the maximum of the quantum noise with equation \eref{ww} (or similar ones depending on the condition of the experiment) will, in fact, conclude that in order to extract all the phenomenology in scattering phenomena, the time-dependent evolution and localized nature of the electron cannot be neglected.

\section*{Acknowledgement}
This work has been partially supported by the \lq\lq{}Ministerio de Ciencia e Innovaci\'{o}n\rq\rq{} through the Spanish Project TEC2012-31330 and by the Grant agreement no: 604391 of the Flagship initiative  \lq\lq{}Graphene-Based Revolutions in ICT and Beyond\rq\rq{}.

\end{document}